\documentclass[fleqn,usenatbib]{mnras}
\usepackage{natbib}
\usepackage{newtxtext,newtxmath}
\usepackage{orcidlink}

\DeclareRobustCommand{\VAN}[3]{#2}
\let\VANthebibliography\thebibliography
\def\thebibliography{\DeclareRobustCommand{\VAN}[3]{##3}\VANthebibliography}

\usepackage[T1]{fontenc}
\usepackage{graphicx}
\usepackage{xcolor}
\newcommand{\orcid}[1]{\href{https://orcid.org/#1}{\textcolor[HTML]{A6CE39}{\aiOrcid}}}
\DeclareUnicodeCharacter{2212}{-}

\usepackage{graphicx}	
\usepackage{amsmath}	

\title[Quasar absorption outflows on galactic scales]{Quasar absorption outflows on galactic scales:\\ Insights from DESI}

\author[M. Dehghanian et al.]{
Maryam. Dehghanian,$^{1}$\orcidlink{0000-0002-0964-7500}\thanks{E-mail: m.dehghanian@uky.edu}
Nahum. Arav,$^{2}$ \orcidlink{0000-0003-2991-4618}
Mayank Sharma,$^{2}$\orcidlink{0009-0001-5990-5790}
Gwen Walker,$^{2}$\orcidlink{0000-0001-6421-2449}
\newauthor Kyle. Johnston,$^{2}$
Matthew. Kaupin,$^{2}$
Justin. Gandhi,$^{2}$
\\
$^{1}$Department of Physics and Astronomy, The University of Kentucky, Lexington, KY 40506, USA\\
$^{2}$Department of Physics, Virginia Tech, Blacksburg, VA 24061, USA}

\date{Accepted 2025 May 27. Received 2025 May 23; in original form 2025 February 17}

\pubyear{\the\year{}}

\begin{document}
\label{firstpage}
\pagerange{\pageref{firstpage}--\pageref{lastpage}}
\maketitle

\begin{abstract}
Absorption outflows in quasars play an important role in understanding active galactic nuclei (AGN) feedback and their influence on galaxy evolution. The unprecedented spectral data provided by the Dark Energy Spectroscopic Instrument (DESI) opens new avenues to explore these outflows. 
We analyze five low-ionization absorption outflow systems in four intermediate-redshift quasars ($2 < z < 3$) using the data obtained by DESI in order to characterize their physical properties and energetics, and also to assess their role in AGN feedback.
We use the spectra from DESI's Early Data Release to determine the ionic column densities, total hydrogen column densities, electron number densities, and ionization parameters via photoionization modeling and absorption line analysis. We derive the outflows' distance from the AGN, and its kinematic properties: mass-flow rates, kinetic luminosity, and momentum flux. 
Our study identifies five distinct mini-broad absorption line outflow systems, hosted by four quasars. The identified outflows exhibit hydrogen column densities of $\log (N_H) = 20.0-20.7$[cm$^{-2}$], ionization parameters of $\log (U_H) = -2.7$ to $-2.1$, and electron number densities of $\log (n_e) =1.45-2.85$[cm$^{-2}$]. The distances of the outflows from the central source are between 4.5 to 31 kpc, and the kinetic luminosities range from $2\times10^{-6}$ to $5\times10^{-3}$ of the Eddington luminosity. The outflows in J1407 and J1032 show the largest kinetic luminosities, with momentum flux ratios ($\dot{p} / \dot{p}_{rad}$) of about 2 and 0.2, respectively. Our findings highlight the vital role of DESI data in uncovering the diversity and significance of quasar outflows in galaxy evolution.
\end{abstract}




\section{Introduction}

Absorption outflows in quasars provide insights into the evolution of the active galactic nuclei (AGNs) and their feedback processes. Understanding these outflows, which are identified through blue-shifted absorption lines in quasar spectra, is vital for unraveling their role in regulating star formation, enriching the intergalactic medium with metals, and driving feedback mechanisms that influence galaxy evolution \citep{king15, some15}.
Quasar absorption lines associated with outflows are classified into three primary categories based on their width: 

\noindent 1. Broad absorption lines (BALs) are characterized by \ion{C}{iv}~$\lambda\lambda 1548,1551$\AA\  troughs with widths exceeding 2,000 km~s$^{-1}$ \citep{weym91,chur99,ham04,itoh20,ham13}.‘LoBALs’ are a subset of BALs that in addition to the high-ionization BALs (e.g., \ion{C}{iv}~$\lambda\lambda 1548,1551$\AA\ and \ion{Si}{iv}~$\lambda\lambda 1394,1403$\AA), show low-ionization absorption lines such as \ion{Mg}{ii} $\lambda\lambda$2798,2803\AA, and \ion{Al}{iii} $\lambda\lambda$1855,1863 \citep[e.g.,][and refs. therein]{ham19,trump06}. 

\noindent 2. Mini-BALs, have velocities widths between 500 and 2000 km~s$^{-1}$ \citep [e.g.,][]{gang08,deh24,deh24b,deh25b}. 

\noindent 3. Narrow absorption lines (NALs) exhibit widths less than 500 hundred km~s$^{-1}$ \citep[e.g.,][]{misa07,deh24storm}.    

Quasar outflows are also identified from blue-shifted emission lines, particularly high-ionization broad lines such as \ion{C}{iv}, which are often offset by several hundred to thousands of km s$^{-1}$ from systemic \citep[e.g.,][]{rich11,coat16}. These emission-line shifts are believed to originate in outflowing gas from the broad-line region and serve as an additional diagnostic of AGN-driven winds.

The Dark Energy Spectroscopic Instrument (DESI) marks a transformative advance in spectroscopic surveys, delivering unprecedented amount of quasar spectra. DESI’s ability to observe tens of millions of galaxies and quasars across a broad wavelength range has provided a rich dataset for studying absorption features. As of June 2023, DESI has identified approximately 90,000 quasars in its Early Data Release \citep[EDR,][]{DESI}, which accounts for about 2 per cent of the survey's ultimate goal (from DESI lab\footnote{https://www.desi.lbl.gov/2023/06/13/the-desi-early-data-release-in-now-available/}). The EDR provides resource for the scientific community, enabling assessments of DESI’s data quality, calibration processes, and its potential for conducting cosmological studies.

In this study, we use the EDR data of four
intermediate-redshift quasars ($2<z<3$) to identify absorption systems showing troughs of excited states of \ion{Si}{ii} and/or \ion{C}{ii}. In section 3.5 we show that the physical origin of these systems are most probably outflows emanating from the quasar. Our analysis includes: 


\begin{enumerate}
    \item Measure the ionic column densities ($N_{\textrm{ion}}$) of the outflow's troughs.
    \item Derive constraints on the physical and kinematic properties of the outflows, including the total hydrogen column density ($N_{\textrm{H}}$), ionization parameter ($U_{\textrm{H}}$), electron number density ($n_{\textrm{e}}$), and distance from the AGN ($R$).
    \item Determine the energetics of the outflow, including the mass-flow rate and kinetic luminosity, to assess its possible role in the AGN feedback processes.
\end{enumerate}

This paper is structured as follows: 
In Section~\ref{obs}, we explain the observations and show one example of the spectral data, including the relevant absorption lines. Section~\ref{sec:anal} includes detailed approaches to measure $N_{\textrm{ion}}$, the results of photoionization modeling and also the details of $n_{\textrm{e}}$ measurements. Section~\ref{sec:anal} also describes how we determine the outflows' energetics, and ends with notes on the individual objects. Section~\ref{sec:disc} provides a summary of the paper.

Here we adopted standard $\Lambda$CDM cosmology with h= 0.677, $\Omega_{m}$= 0.310, and
$\Omega_\Lambda$ = 0.690 \citep{plan20}. We used the Python astronomy
package Astropy \citep{astro13,astro18} for our
cosmological calculations, as well as Scipy \citep{virt20},
Numpy \citep{harr20}, and Pandas \citep{reba21} for most of our numerical computations. For our plotting purposes, we used Matplotlib \citep{hunt07}.
\begin{figure*}
\includegraphics[width=7 in]{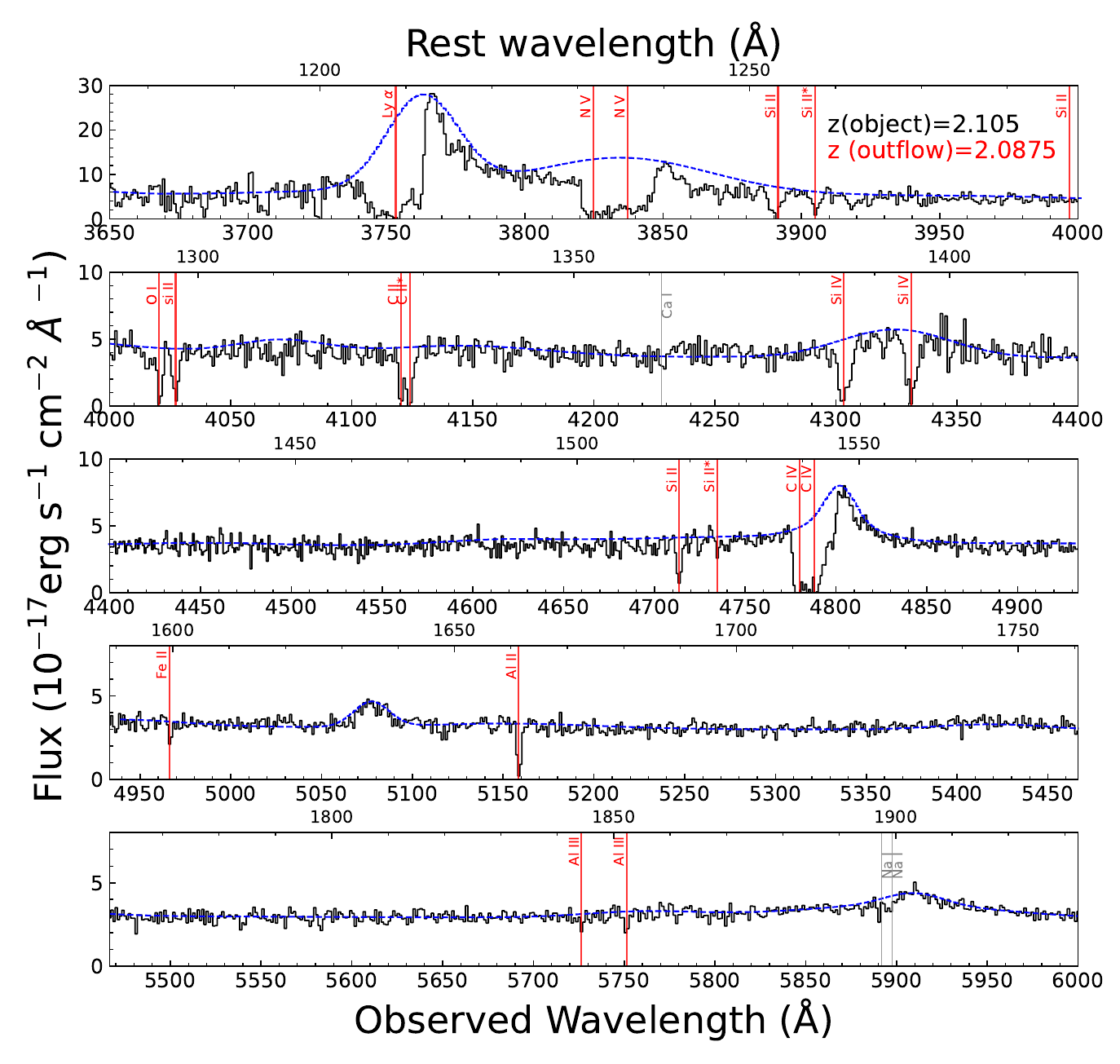}
\caption{The spectrum of SDSS J0831+3340 as observed by the DESI in 2021. The absorption features of the outflow system with a velocity of $-$1700 km s$^{-1}$ are marked with red lines. The dashed blue line shows our continuum emission model.}
            \label{figFlux1}%
\end{figure*}
\section{Sample selection and observations}
\label{obs}

For our sample selection, we utilize the value-added catalogs (VACs) produced by the DESI
science collaboration \citep{fil24}. 
We are interested in objects that show absorption troughs from both ground and excited states of either (or both) \ion{Si}{ii} and \ion{C}{ii}, as such features allow us to get the distance of the outflow from the quasar.  These transitions are found in 
the rest-frame wavelength
range of 1250–1550 \AA  To detect these features
we restrict our sample to quasars with redshifts greater than 2, ensuring
that the relevant spectral region is accessible within the observed
frame of the DESI observations. To maintain a sufficient signal-to-noise
ratio for our analysis, we select only quasars with g-band magnitudes
brighter than 21.4. Applying these criteria, we obtain our sample of
four quasars. 

The selected quasars were observed by DESI, covering
a spectral range of 3600–9800 \AA\ with a resolution of R$\approx$2000-5000,
depending on wavelength. All observations were conducted during
the sv1 survey validation phase and processed using the DESI data
pipeline. Details of the individual objects and their observations are
provided in Table~\ref{targets}. Hereafter, we refer to each object by the first
four digits of its RA.

\subsection{The Redshift of the objects}
The quasar spectra in DESI-EDR are accompanied by the redshifts determined by the DESI pipeline \citep[described in][]{chas23}. During our visual inspection, we find that in many cases, the location of major emission lines (e.g., \ion{C}{iv} and \ion{Si}{iv}) predicted by the pipeline is significantly offset from the actual data (by up to $\sim$ 500 - 1000 km~s$^{-1}$).  \cite{wu23} investigate this issue for all quasars in the DESI-EDR and provide improved systemic redshifts by obtaining fits from multiple emission lines and calculating their weighted average. For our sample, we find that their redshift values (reported in Table~\ref{targets} along with their uncertainties) provide a much better visual fit to the emission lines and we thus use them for our analysis.


\label{sec:obs}
\begin{table*}
      
        \centering
\begin{tabular}{ccccccc}
\hline
\noalign{\smallskip}
Target name &  Redshift & RA & DEC&epoch 1&epoch 2&Total exposure time (s) \\
\noalign{\smallskip}
\hline
\noalign{\smallskip}
J0831+3340  & 2.0993$\pm$0.0029&08:31:26.06&+33:40:37.77&2021-03-20&2021-04-03&4258 \\
J1032+3111 & 2.6051$\pm$0.0048&10:32:30.98&+31:11:43.81&2021-01-10&2021-02-09&7830 \\
J1407+5110 & 2.6034$\pm$0.0034&14:07:29.90& +51:10:23.63&2021-01-17&2021-04-30&24359\\
J1609+1516&2.1208$\pm$0.0030&16:09:51.96& +15:16:03.71&2021-03-19&2021-04-05&4532\\
\noalign{\smallskip}
\hline
\end{tabular}
\caption{\label{targets}Targets of this study}
\end{table*}

\section{Analysis}
\label{sec:anal}
We show all the details of our analysis for one quasar (J0831) and perform the same analysis on the other objects.
The spectrum of quasar J0831 is shown in Fig.~\ref{figFlux1}. This figure also shows the identified absorption troughs with vertical red lines. We normalize the spectrum by dividing the observed data by the unabsorbed emission model. The unabsorbed emission model is constructed using a single power law, while one or more Gaussians reproduce the emission features. 

\begin{figure}
\includegraphics[width=\columnwidth]{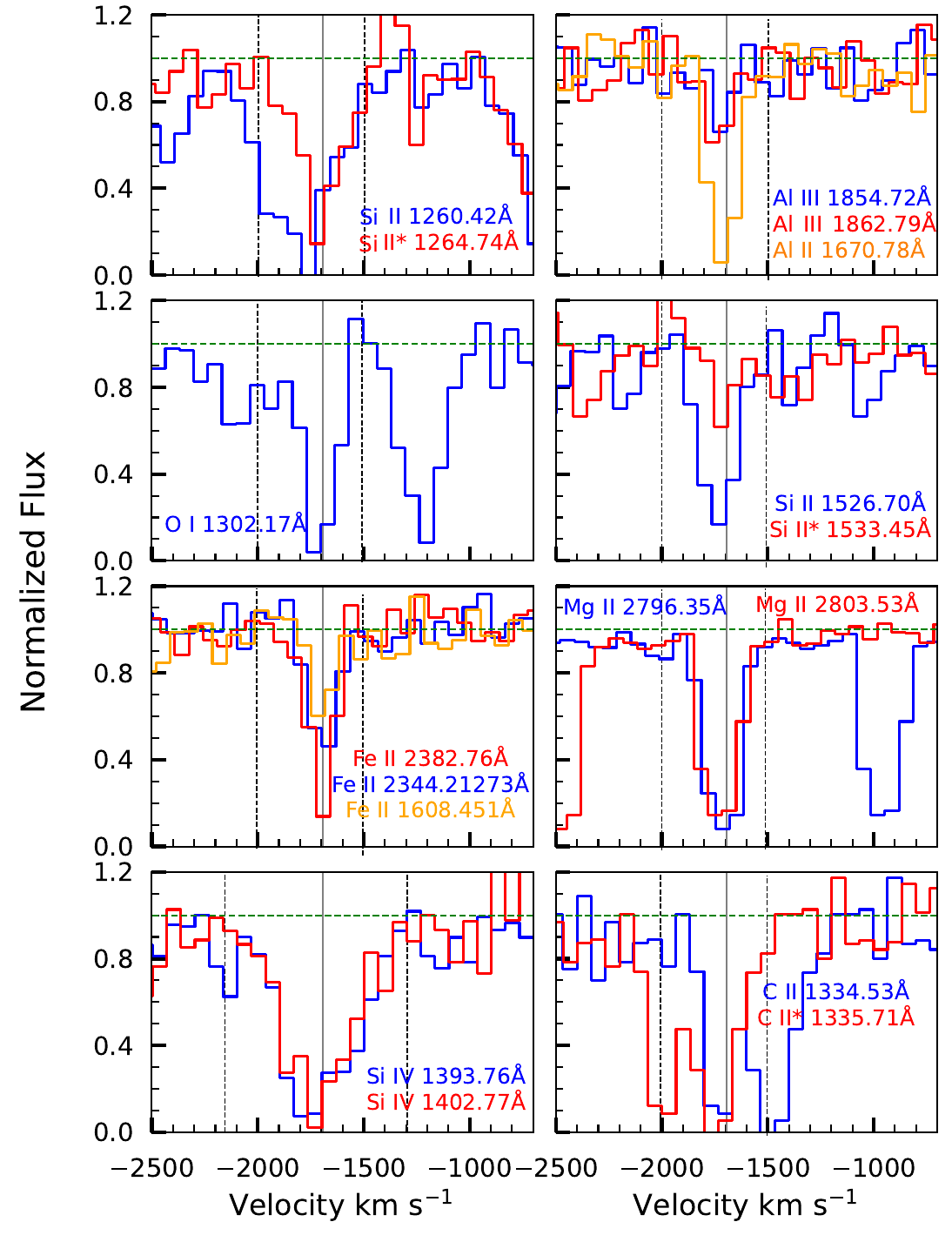}
\caption{Normalized flux versus velocity for outflow's absorption troughs detected in the spectrum of J0831. The horizontal green dashed line shows the continuum level,
and the vertical black dashed lines show the integration range (see text). The vertical solid gray line 
indicates the centroid velocity.}
            \label{figvelocity}%
\end{figure}

Fig.~\ref{figvelocity} displays the normalized flux as a function of velocity for various blue-shifted absorption lines detected in the spectrum of J0831. Although the figure includes both low- and high-ionization absorption troughs, this paper focuses on the $N_{\textrm{ion}}$ of  \ion{Al}{ii}, \ion{Al}{iii}, \ion{Si}{ii}, \ion{Si}{ii*}, \ion{C}{ii}, and \ion{C}{ii*}. This selection is motivated by the need to determine $N_{\textrm{H}}$, $U_{\textrm{H}}$, and $n_{\textrm{e}}$ accurately. Using pairs of ions from the same element, such as \ion{Al}{ii} and \ion{Al}{iii}, minimizes dependence on elemental abundances, thereby improving the reliability of the derived parameters \citep{deh24b}. We use the ratio of the excited states to the ground state of \ion{Si}{ii} for $n_{\textrm{e}}$ determination (see Section~\ref{sec:ne}). Fig.~\ref{figvelocity} also shows the velocity integration range we considered in our column density calculations (dashed vertical lines). In the case of J0831, the velocity range considered for the low-ionization lines is between $-2000$ to $-1500$ km~s$^{-1}$. 

\begin{table*}
\begin{tabular}{llllllll}
\hline
\noalign{\smallskip}
Ion&$\lambda_{\textrm{rest}}$(\AA)&J0831& J1032& J1407,S1& J1407,S2& J1609 \\
\noalign{\smallskip}
\hline
\noalign{\smallskip}
\ion{Al}{ii} &1670.78&$42^{+7.7}_{-11.0}$&$4.9^{+1.6}_{-1.4}$&$9.6^{+1.8}_{-1.6}$&19$^{+2.8}_{-2.5}$&$29^{+6.8}_{-4.7}$\\\noalign{\smallskip}
\ion{Al}{iii} &1854.72,1862.79&$61^{+16}_{-19}$&$25^{+7.4}_{-6.5}$&$25^{+7.4}_{-6.7}$&35$^{+7.6}_{-6.8}$&$110^{+59}_{-28}$\\\noalign{\smallskip}
\ion{C}{ii}  &1334.53&$800^{+230}_{-230}$&$380^{+120}_{-120
}$&$570^{+61}_{-48}$&$830^{+180}_{-95}$&$1400^{+600}_{-300}$ \\\noalign{\smallskip}
\ion{C}{ii*}&1335.71&$1200.0^{+350}_{-350}$&$450^{+180}_{-180}$&$580^{+54}_{-42}$&$1240^{+420}_{-180}$&$650^{+330}_{-120}$\\\noalign{\smallskip}
\ion{Fe}{ii} &1608.45&$220^{+59}_{-72}$&--&--&--&$120^{+30}_{-23}$\\\noalign{\smallskip}
\ion{Si}{ii} &1260.42&$>145_{-30}$&$>35_{-4}$&$>70_{-5}$&$>90_{-10}$&$>70_{-10}$\\\noalign{\smallskip}
\ion{Si}{ii*} &1264.74&$>69_{-14}$&$>40_{-5}$&$>41_{-4}$&12$^{+2.7}_{-2.5}$&$>50_{-8}$\\\noalign{\smallskip}
\ion{Si}{ii} &1526.70&$530^{+140}_{-71}$&$260^{+31}_{-27}$&$140^{+20}_{-18}$&350$^{+34}_{-28}$&$460^{+77}_{-48}$\\\noalign{\smallskip}
\ion{Si}{ii*} &1533.45&$120^{+38.}_{-30}$&$80^{+19}_{-16}$&$80^{+16}_{-19}$&--&$100^{+18}_{-16}$\\\noalign{\smallskip}
\hline
\end{tabular}
\caption{\label{tab3}Measured Column Densities. The ionic column densities of the absorption 
lines detected in the outflows. For each 
measurement, a 10 per cent systematic error is 
quadratically added to the error arising from
the measurement (see \protect\cite{deh24b}). All column 
density values are in units of 10$^{12}$cm$^{-2}$.
}
\end{table*}
\begin{figure}
\includegraphics[width=\columnwidth]{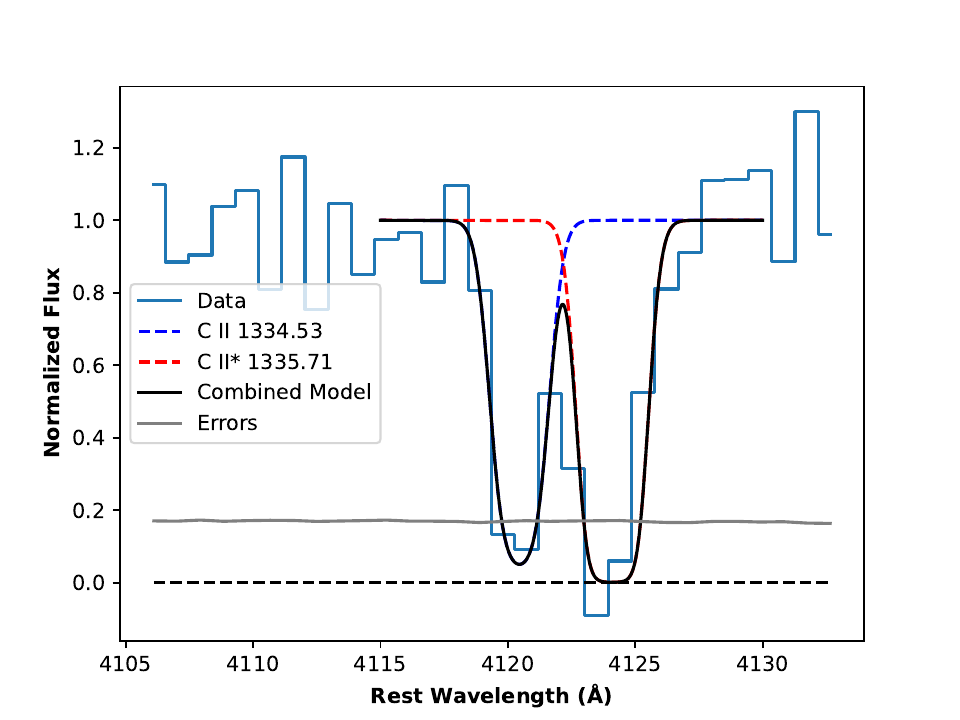}
\caption{Gaussian modeling of the \ion{C}{ii*}$ \lambda$1335.71 \AA\ (red dashed line) and \ion{C}{ii} $\lambda$1334.53
\AA\ (blue dashed line) absorption troughs in the absorption outflow of J0831. The black solid line shows the final model, which results from combining the two Gaussian curves. The grey line shows the level of noise around the modeled region.}
            \label{figC2}
\end{figure}
\subsection{ionic column densities}
For each quasar, we determine $N_{\textrm{ion}}$ of the identified absorption troughs as the first step to extract the physical characteristics of the outflow. To do so, we either use the "apparent optical depth" (AOD) method or the "partial coverage" (PC) method. Details about each approach are as follows:

\noindent 1. When using the AOD method, we assume the outflow uniformly covers the whole source. 
    Using this assumption, we determine the column densities using \citep[e.g.,][]{spit68, sava91,arav01}: 
\begin{equation}
I(v)=e^{-\tau(v)}, \label{eq-1}
\end{equation}
\begin{equation}
N_{\textrm{ion}}=\frac{3.77\times 10^{14}}{\lambda_{0}f}\times \int \tau(v)~dv 
 [\textrm{cm}^{-2}],
\label{eq0}
\end{equation}
\noindent in which $I(v)$ is the normalized intensity profile as a function of velocity and $\tau(v)$ is the optical depth of the absorption trough. $\lambda_{0}$ is the transition's wavelength, $f$ is the oscillator strength of the transition, and $v$ is measured in km~s$^{-1}$. Here, the AOD method is used whenever we are dealing with a singlet, such as \ion{Al}{ii}, \ion{Si}{ii}, and \ion{Si}{ii*} \citep[e.g.,][]{arav01}.

\noindent2. The PC method is used to calculate the ionic column densities of the \ion{Al}{iii} doublet, in which two lines originate from the same energy level. The main assumption in this method is that the outflow partially covers a homogeneous source \citep{barl97, arav99a, arav99b} and its coverage and optical depths are calculated using:
\begin{eqnarray}
    I_R(v)-[1-C(v)]=C(v)e^{-\tau(v)} \label{eq30}
\end{eqnarray}
and
\begin{eqnarray}
    I_B(v)-[1-C(v)]=C(v)e^{-2\tau(v)}.\label{eq40}
\end{eqnarray}
\noindent These equations assume doublet transitions where the blue component has twice the oscillator strength of the red component, and $I_R(v)$ and $I_B(v)$ are the normalized intensities of the red and blue doublet lines, respectively, while $\tau(v)$ is the optical depth profile of the red component. Note that considering a velocity-dependent covering factor assures that the effects of non-black saturation \citep{kool02} are included. We insert the calculated $C(v)$  and $\tau(v)$ in the PC method equations to calculate the column density of \ion{Al}{iii} ion.
See \cite{barl97,arav99a,arav99b, kool02, arav05,borg12b,byun22b, byun22c} to find detailed explanations of both methods.    

For all five outflows, absorption troughs from \ion{C}{ii} 1334.53 \AA\  and \ion{C}{ii*} 1335.71 \AA \ were identified. Due to the small separation between these two troughs, each was first modeled using a Gaussian profile to disentangle their contributions. Subsequently, the AOD method was employed to calculate their respective column densities. Figure~\ref{figC2} illustrates this approach for the troughs identified in J0831. 
\subsection{Photoionization modeling}
\label{sec:photo}
The $N_{\textrm{ion}}$ are used as the starting point for deriving $N_{\textrm{H}}$ along the line of sight and $U_{\textrm{H}}$, defined as the ratio of ionizing photon density to hydrogen number density, given by \citep{oster06}:
\begin{equation}
U_\textrm{H}\equiv\frac{Q_{\textrm{H}}}{4\pi R^{2}c~n_{\textrm{H}}} \label{eq-U}
\end{equation}
\noindent where $Q_{\textrm{H}}$ is the number of hydrogen-ionizing photons emitted per second by the source; $R$ is the distance from the source to the absorbing gas; $n_{\textrm{H}}$ is the hydrogen number density, and $c$ is the speed of light.

Using the observed $N_{\textrm{ion}}$, we constrain $N_{\textrm{H}}$ and $U_{\textrm{H}}$ to characterize the physical conditions of the gas. This analysis employs the spectral synthesis code Cloudy \citep{cloud23}, which solves the ionization equilibrium equations for various physical conditions.

The outflow is modeled as a plane-parallel slab with a constant $n_{\textrm{H}}$ 
and solar metallicity \citep{grev98,hol01}. The spectral energy distribution (SED) incident upon the outflow affects its photoionization state. To remain consistent with \cite{xu19} who conducted a similar survey of quasar outflows, we use the UV-soft \citep{dun10,arav13} SED as the input for our Cloudy calculations.

\begin{figure}
\includegraphics[width=\columnwidth]{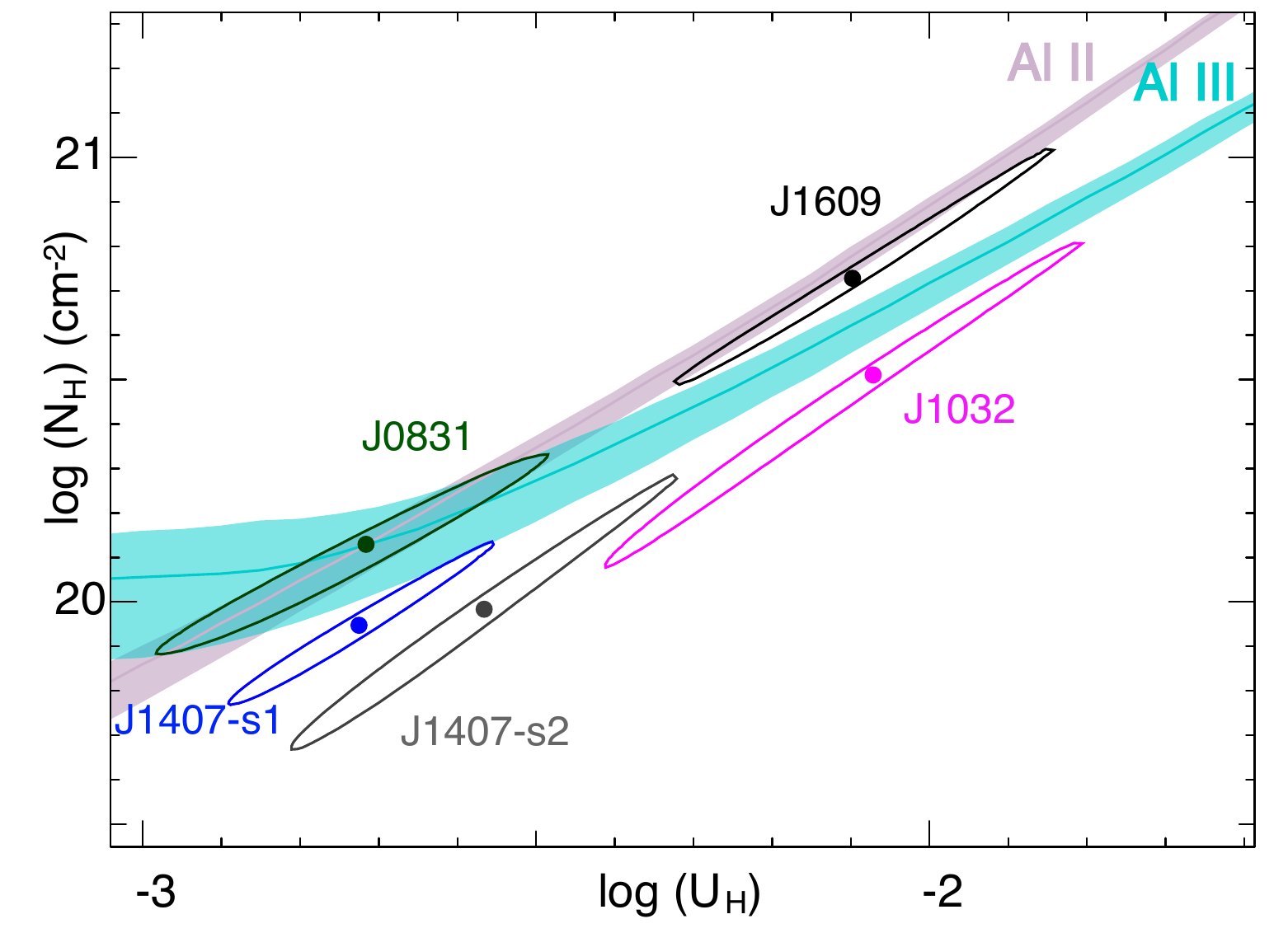}
\caption{Photoionization solution for six outflow systems detected in our five quasars. Each colored band indicates the $N_{\textrm{H}}$ and $U_{\textrm{H}}$ region consistent with the measurement (solid line) of J0831 and their errors (shaded region), assuming the UV-soft SED and solar abundances. For each quasar, the colored dot shows the best solution that reproduces the observed \ion{Al}{ii} and \ion{Al}{iii} ions, while an oval of the same color shows the 1$\sigma$ uncertainty.}
            \label{figNVU}%
\end{figure}
Cloudy generates a grid of models within the ($N_{\textrm{H}}$,$U_{\textrm{H}}$) parameter space. Each grid point predicts the total column densities for all relevant ions. By comparing the measured $N_{\textrm{ion}}$ with Cloudy predictions, we identify solutions that satisfy the observational constraints. This is visualized in figure~\ref{figNVU} as a phase-space plot showing the photoionization equilibrium solutions for the observed ions. In this figure, each dot shows the best-fit model for the quasar, corresponding to a pair ($N_{\textrm{H}}$, $U_{\textrm{H}}$) surrounded by an oval of the same color that shows the 1-$\sigma$ deviation. The best-fit model is determined by minimizing the \(\chi^2\) statistic. Note that in figure~\ref{figNVU}, the \ion{Al}{ii} and \ion{Al}{iii} contours show the measured values (solid lines) and their uncertainties (the shaded bands) calculated for the quasar J0831. All ($N_{\textrm{H}}$, $U_{\textrm{H}}$) values resulting from the photoionization modeling are displayed in columns 2 $\&$ 3 of Table~\ref{table:outflowpar}.

\subsubsection{Sensitivity to the assumed SED, abundances  and constant nH }
(a) Considering other commonly used quasar SEDs such as HE0238 \citep{arav13} or MF87 \citep{mat87}
changes the derived values of both NH and UH by$ \sim \pm 0.2$ dex  \cite[e.g., section 3.2 and especially Fig 5 from][]{deh25b}.
(b) Changing the assumed abundances  also affect these derived values.  For the extreme case of 4.6 times solar metalicity, $N_{\textrm{H}}$ is reduced by about an order of magnitude and $U_{\textrm{H}}$ is reduced by about 0.5 dex. \cite[see again section 3.2 and especially Fig 5 from][]{deh25b}.(c) The assumption of constant $n_{\textrm{H}}$ is empirically justified by high-resolution and high signal-to-noise spectra of quasar outflows.  For example, using 
Echelle spectra (taken by VLT/UVES), figure 2 of \cite{kor08} shows that the column density ratio of \ion{Fe}{ii*} to \ion{Fe}{ii} stays roughly constant over many resolution elements of that outflow's troughs. The value of such a ratio yields the electron number density of the 
gas and therefore its $n_{\textrm{H}}$ (see section 3.3 here). Thus, this results supports the assertion 
that $n_{\textrm{H}}$ is constant across the kinematic profile of the outflow.

\begin{table*}
\setlength{\tabcolsep}{6pt}
\renewcommand{\arraystretch}{1.33}
\centering
\begin{tabular}{lccccccccc}
\hline
 &(1)&(2)&(3)&(4)&(5)&(6)&(7)&(8)&(9)\\
Object&$v_{\textrm{centroid}}$& $\textrm{log}(U_\textrm{H})$& $\textrm{log}(N_\textrm{H})$& $\textrm{log}(n_\textrm{e})$&\textrm{R}&$\Dot{M}$&$\dot{p}/\dot{p}_{\textrm{rad}}$&$\log (\Dot{E_{\textrm{K}}})$&$\Dot{E_{\textrm{K}}}/{L_{\textrm{Edd.}}}$ \\
&(km~s$^{-1}$)&[dex] & [$\textrm{cm}^{-2}$] & $[\textrm{cm}^{-3}]$&(kpc)&$(\textrm{M}_{\odot} \textrm{yr}^{-1})$&&[erg s$^{-1}$]& \\
\hline
J0831  &{$-$1100$\pm$280}&$-2.7^{+0.24}_{-0.27}$ &$20.1^{+0.21}_{-0.25}$ &$2.35^{+0.15}_{-0.15}$ &$12^{+5}_{-3.5}$ &{$47^{+15}_{-13}$}&{0.37}&{$43.2^{+0.12}_{-0.14}$}&{$1.5^{+0.5}_{-0.4}\times10^{-3}$}\\
J1032&{$-$1000$\pm$400}&$-2.1^{+0.26}_{-0.33}$&$20.5^{+0.30}_{-0.43}$&$2.5^{+0.15}_{-0.10}$&$6.3^{+4.0}_{-2.3}$&{$58^{+29}_{-27}$}&{0.24}&{$43.3^{+0.18}_{-0.27}$}&{$1.2^{+0.6}_{-0.6}\times 10^{-4}$} \\
J1407,S1 &{$-$3600$\pm$280}&$-2.6^{+0.25}_{-0.24}$&$20.0^{+0.31}_{-0.31}$&$2.85^{+0.15}_{-0.15}$&$5.5^{+2.1}_{-1.6}$&{$58^{+34}_{-22}$}&{1.7}&{$44.4^{+0.20}_{-0.20}$}&{$4.7^{+2.7}_{-1.7}\times 10^{-3}$}\\
J1407,S2 &{$-$700$\pm$285}&$-2.7^{+0.17}_{-0.17}$&$20.0^{+0.19}_{-0.18}$&$1.45^{+0.1}_{-0.1}$&$31^{+7.6}_{-6.4}$&{$63^{+19}_{-14}$}&{0.36}&{$43.0^{+0.12}_{-0.11}$}&{$1.9^{+0.6}_{-0.4}\times 10^{-4}$}\\
J1609&{$-$230$\pm$290}&$-2.1^{+0.25}_{-0.23}$&$20.7^{+0.29}_{-0.23}$&$2.35^{+0.1}_{-0.1}$&$4.5^{+1.5}_{-1.1}$&{$15^{+7}_{-4}$}&{0.04}&{$41.4^{+0.17}_{-0.12}$}&{$1.8^{+0.9}_{-0.5}\times10^{-6}$}\\
\hline
\end{tabular}
\caption{DESI survey: Outflow Parameters.\\The values derived in columns (2), (3) and (4) are based on using solar metalicity and the UV-soft SED.  The systemic errors that arise from using different metalicity and SEDs are discussed in section 3.2.1. The values derived in columns (5), (6), (7), (8) and (9) are based on the derived parameters in  columns (2), (3) and (4) and their errors, and do not include the systemic velocity errors (given in column (1)) associated with the redshift determination of each quasar (see Table 1). Column (7) gives the ratio of the outflow's momentum flow rate, to the momentum flux of associated with the quasar's luminosity (L/c).}
\label{table:outflowpar}
\end{table*}
\subsection{Electron number density}
\label{sec:ne}
\begin{figure}
\includegraphics[width=\columnwidth]{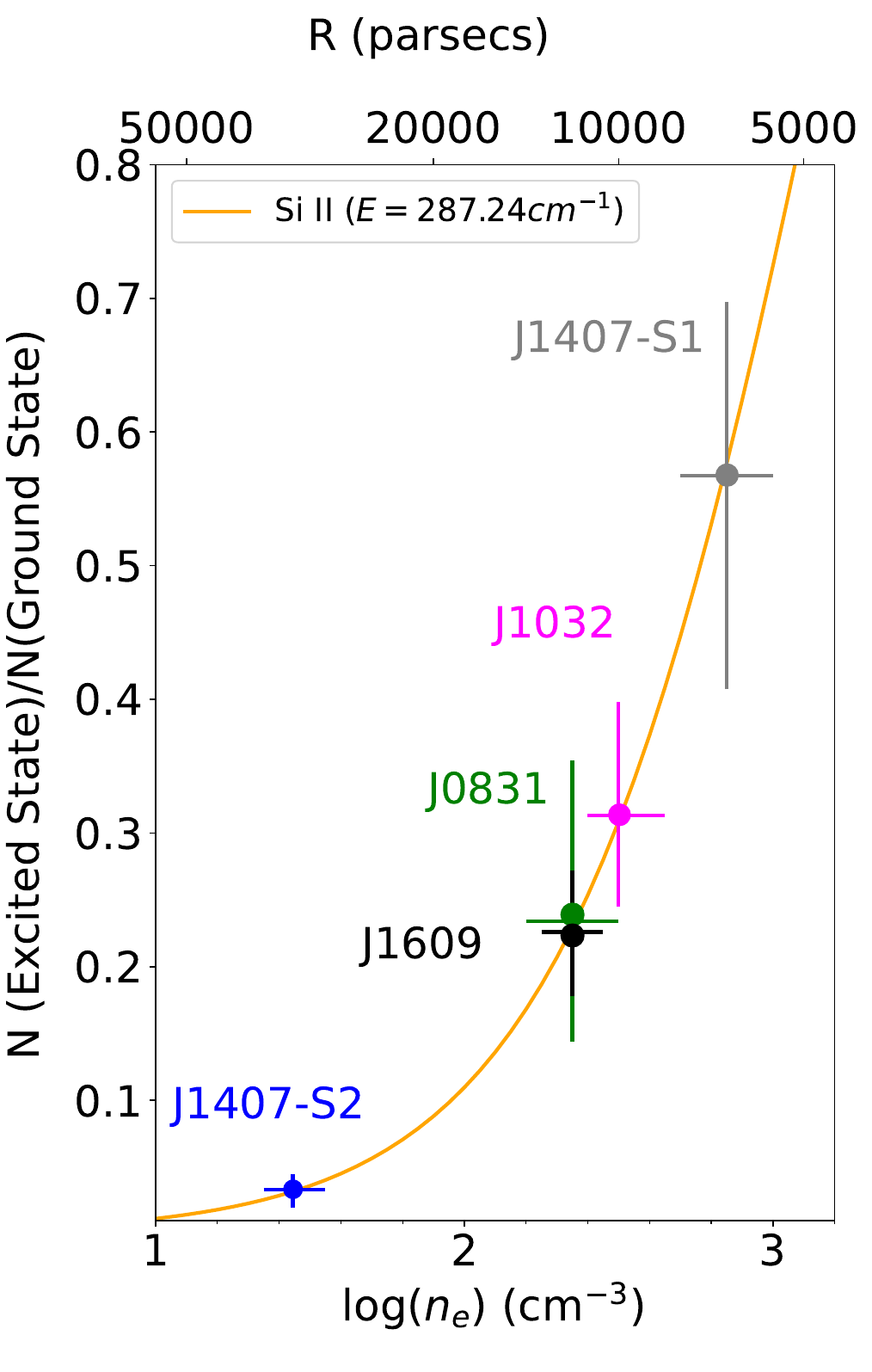}
\caption{Excited state to resonance state column density ratio of \ion{Si}{ii} (orange curve) vs. the electron number density from the Chianti atomic database for T=10000K. A colored dot shows the measured ratio for each outflow system. The 287.24 cm$^{-1}$ transition in \ion{Si}{ii} arises from the $^2P_{3/2} \, \rightarrow \, ^2P_{1/2}$  transition within this ion.}
            \label{figchianti}
\end{figure}

Chianti \citep{dere97} is an atomic database designed for modeling emission and absorption spectra of ionized plasmas in astrophysical environments. Here, we use its version 9 \citep{dere97, dere19} to calculate n$_{e}$ of the outflows, using the abundance ratios of the excited state to the resonance state for \ion{Si}{ii}. 
For a particular temperature ($\approx 10^{4} \textrm{K}$ for our outflows as predicted by the Cloudy's solution), Chianti calculates the column density ratio of excited to ground state as a function of $n_{\textrm{e}}$, using the atomic data for the desired ion, including energy levels, radiative transition rates, and collisional excitation rates. We compare Chianti's prediction with the observations (our $N_{ion}$ measurements) to estimate $n_{\textrm{e}}$. Figure~\ref{figchianti} displays the results.
To obtain the uncertainties for $n_{\textrm{e}}$, we followed the methodology of \cite{deh24b}.

Figure~\ref{figchianti} also shows the distance between the outflow and the central source, as its top horizontal axis. While the values on the axis belong to the J0831 outflow, the same logic works for all outflows: the location is derived from equation(\ref{eq-U}), which can be written as: 
\begin{equation}
R=\sqrt{\frac{Q_{\textrm{H}}}{4\pi c~n_{\textrm{H}} U_{\textrm{H}}}} \label{eq-R}
\end{equation}
\noindent in which $U_{\textrm{H}}$ is taken from the photoionization solution and  $n_{\textrm{e}}\approx1.2~n_{\textrm{H}}$ \citep[][for a highly ionized plasma]
{oster06}. Note that this relation could be different for LoBALs (see section 4.3 of \cite{sharma25}; however, we confirm with our Cloudy models that this was not the case for any of our outflows. 

For each quasar, we calculate the value of $Q_{\textrm{H}}$ via scaling the UV-soft SED with the observed continuum flux of each quasar at $\lambda_{\textrm{observed}}$ between 4500 to 5500 \AA\  (depending on the quasar's redshift it will be between 1250 to 1750 \AA\ in the rest frame). We then use the quasar's redshift, adopted \(\Lambda\)CDM cosmology, and the scaled SED to obtain the bolometric luminosity and $Q(\textrm{H})$  \citep [see][]{byun22a,byun22b,walk22}. All resulted values of $n_{\textrm{e}}$ and $R$ are available in Table~\ref{table:outflowpar}, columns 4 $\&$ 5, respectively.

\subsection{Outflow's energetics}
\label{sec:enrg}


The energetics of the outflow, i.e., its mass-flow rate, momentum-flux rate, and kinetic luminosity, are calculated using the equations below \citep{borg12b}: 
\begin{equation}
\Dot{M} \simeq 4\pi \Omega R N_\textrm{H} \mu m_{\textrm{p}} \nu \label{eq5}
\end{equation}
\begin{equation}
{\Dot{p}= \Dot{M} v\label{eqpdot}}
\end{equation}
\begin{equation}
\Dot{E}_{\textrm{K}} \simeq \frac{1}{2}\Dot{M} v^{2}\label{eq6}
\end{equation}
\noindent where $\mu$ = 1.4 is the mean atomic
mass per proton, $v$ is the outflow's velocity, and m$_{p}$ is the proton's mass. $\Omega$ is the global covering factor defined as the fraction of the solid angle around 
the source covered by the outflow. We assume $\Omega \approx$0.2 based on the ratio of quasars that show a \ion{C}{iv} BAL(e.g., 
\citealt{hew03,dai12,gib09,all11}). 
To determine the BH mass and its Eddington luminosity, we follow the discussion in \cite{deh24} and references therein: we measure the FWHM of \ion{C}{iv} emission line for each case, and then use its blue-shift, $v_{\textrm{centroid}}$, and flux to measure the BH mass and Eddington luminosity. The results of these calculations for all five quasars are available in Table~\ref{tabMBH}. Note that since we use the corrected formulas that consider the \ion{C}{iv} blue-shift, the large SED-dependent systematic error in BH masses is removed \citep{coat17}. 

\begin{table*}      
     $$
\begin{tabular}{lcccc}
\hline
\noalign{\smallskip}
Object&\ion{C}{iv} FWHM& \ion{C}{iv} Blueshift& $M_{\textrm{BH}}$& $\textrm{log}(L_{\textrm{Edd.}})$ \\
&(km~s$^{-1}$)&(km~s$^{-1}$)&($\times10^{8}\textrm{M}_{\odot}$)& [erg s$^{-1}$] \\
\noalign{\smallskip}
\hline
\noalign{\smallskip}
J0831+3340  &1550&380&$1.01^{+0.15}_{-0.10}$&$46.07^{+0.16}_{-0.11}$\\\noalign{\smallskip}
J1032+3111 &4250&100&$12^{+1.7}_{-1.4}$&$47.18^{+0.16}_{0.15}$ \\\noalign{\smallskip}
J1407+5110 &2770&80&$3.7^{+0.32}_{0.23}$&$46.7^{+0.10}_{0.15}$\\\noalign{\smallskip}
J1609+1516&5900&400&$10.5^{+0.14}_{-0.12}$&$47.13^{+0.16}_{-0.13}$\\\noalign{\smallskip}
\noalign{\smallskip}
\hline
\end{tabular}
     $$ 
\caption{\label{tabMBH}DESI Survey: quasars' Properties}
\end{table*}

\cite{hop10} show that when the ratio of $\Dot{E_{\textrm{K}}}/{L_{\textrm{Edd}}}$ exceeds $0.5 \%$, the outflow can have a significant contribution to AGN feedback processes. According to the last column of Table~\ref{table:outflowpar}, the absorption outflow detected in J0831 with a ratio of 0.47 per cent can significantly contribute to feedback processes. 
{\subsection{Physical origin of the five studied absorbers}}

{Several of the detected absorbers exhibit centroid velocities below |2000| km~s$^{-1}$, with respect to the systemic velocity of the quasar rest-frame. This places them in the category of "proximate" or "associated" absorbers \citep{bow14}.  Here we provide evidence that the five absorber systems we analyze here are the product of AGN outflows based on two independent physical
lines of evidence:
\begin{enumerate}
    \item The electron number densities measured in our five absorbers ($\log (n_e)$>1.5 [cm$^{-3}$], see Table\ref{tab3}); are substantially higher than the values typically observed in  circumgalactic absorbers (e.g., $-$4$\leq$$\log (n_e)$ $\leq$$-$2 [cm$^{-3}$], see Fig. 8 in \cite{pro17}), or in the Milky way ISM (e.g., $-$2$\leq$$\log (n_e)$ $\leq$$-$0.3 [cm$^{-3}$], see Table 6 in \cite{red08}). However, our measured $n_e$  are fully consistent with compact, photoionized outflows emenating from the AGN \citep[e.g.,][]{gabel05,borg13}. 
    \item Velocity width of the absorbers.  The wider the absorber, the more difficult it is to associate it with either IGM, CGM or ISM material and hence the higher is the probability that its physical origin is from a dynamical outflow emanating from the quasar.  The narrowest width of a mini-BAL (500 km/s for the widest trough) is usually taken as a strong evidence that the absorber is indeed arising from a quasar outflow \citep[e.g.,][]{misa10}. As we show in section 3.6, all of the absorbing systems studied here satisfy this criterion.
\end{enumerate}}
\subsection{Notes on individual objects}
\label{ind}
All of the five outflow systems have absorption troughs from  \ion{N}{v}~$\lambda\lambda 1239,1243$\AA, \ion{C}{iv}~$\lambda\lambda 1548,1551$\AA, and \ion{Si}{iv}~$\lambda\lambda 1394,1403$\AA. However, all these high ionization absorption troughs are saturated.
\subsubsection{J0831}
The outflow system identified in this quasar has a centroid velocity of $-$1700 km~s$^{-1}$. The wide \ion{N}{v} absorption troughs (with a width of $\sim$2000 km~s$^{-1}$) along with the identified \ion{Fe}{ii} troughs labels it as a FeloBAL. In addition to the High-ionization absorption lines mentioned above and the low-ionization lines discussed earlier, this outflow also {shows absorption from} the \ion{Mg}{ii}~$\lambda\lambda 2796,2803$\AA\  doublet.
\subsubsection{J1032}
The absorption troughs in this outflow have a centroid velocity of $-250$ km~s$^ {-1}$ and a width of {700 km~s$^{-1}$ (based on \ion{N}{v})}, which puts it in the mini-BAL category. 
\subsubsection{J1407}
There are two outflow systems identified in this quasar: S1 {has $v=−$3200 km~s$^{-1}$ and a \ion{C}{iv} width of about 1000 km~s$^{-1}$, designating it as a mini-BAL. S2 has  $v=−$300 km~s$^{-1}$ and a \ion{C}{iv} width of about 700 km~s$^{-1}$, designating it as a mini-BAL as well.}  

\subsubsection{J1609}
Having a {\ion{C}{iv}} width of $\approx${600} km~s$^{-1}$, the outflow system identified in this object is categorized as a {mini-BAL} outflow. This outflow system has a velocity of $-$470 km~s$^{-1}$ and shows absorption troughs from \ion{Mg}{ii}~$\lambda\lambda 2796,2803$\AA, {as well as from \ion{Fe}{ii}}.  
\section{Conclusions}
\label{sec:disc}

In this work, we presented a detailed analysis of galactic-scale absorption outflows in four intermediate-redshift quasars ($2 < z < 3$) observed with DESI. Our study investigated the physical and kinematic properties of five distinct outflow systems. Our key findings include:

\begin{enumerate}
    \item \textbf{Physical properties:} The five outflows have $\log (N_{\textrm{H}})$ ranging from 20.0 to 20.7 [cm$^{-2}$], while their $\log (U_{\textrm{H}})$  span from $-$2.7 to $-2.1$. These results were obtained using photoionization modeling based on $N_{\textrm{ion}}$  measured from DESI spectra. Electron number densities vary significantly between outflows, with {$\log (n_{\textrm{e}}) = 1.45$ to $2.85$} [cm$^{-3}$], calculated using Chianti. 

    \item \textbf{Outflow distances:} The radial distances of the outflows from their central sources span from a few kiloparsecs to tens of kiloparsecs. 
    
    \item \textbf{Energetics:} The kinetic luminosities of the outflows range from {$0.00018\%$ to $0.47\%$} of the Eddington luminosity. Notably, the outflow in J0831 shows a possible significant contributions to AGN feedback, with $\Dot{E_{\textrm{K}}}/{L_{\textrm{Edd.}}}$ ratio of {$\sim0.5\%$}, while its momentum flux ratios ($\dot{p} / \dot{p}_{rad}$) is almost {two.} 
    

    \item \textbf{AGN feedback implications:} {Outflow in J1407} (S1, within the error bars) demonstrates the potential for significant AGN-driven feedback, capable of enriching the circumgalactic medium and quenching star formation. In contrast, smaller kinetic luminosity in {J0831}, J1032, J1609, and J1407-s2 suggests that not all quasars exhibit feedback-efficient outflows.
\end{enumerate}


\section*{Acknowledgements}
MD acknowledges the support from JWST-AR-06419. We also acknowledge support from NSF grant AST 2106249,
as well as NASA STScI grants AR-15786, AR-16600, AR-
16601 and AR-17556.This research uses services or data provided by the SPectra Analysis and Retrievable Catalog Lab (SPARCL) and the Astro Data Lab, which are both part of the Community Science and Data Center (CSDC) program at NSF's National Optical-Infrared Astronomy Research Laboratory. NOIRLab is operated by the Association of Universities for Research in Astronomy (AURA), Inc., under a cooperative agreement with the National Science Foundation. 

\section*{Data Availability}
The EDR of DESI is publicly available at https:// data.desi.lbl.gov/ doc/ where our objects can be queried using their uniquely assigned target ids.

{The full DESI spectra for each target can be viewed at https://www.legacysurvey.org/viewer/desi-spectrum/edr using the RA and Dec coordinates provided.}



\bibliographystyle{mnras}

\begin{thebibliography}{99}







\bibitem[Allen et al.(2011)]{all11} Allen, J.~T., Hewett, P.~C., Maddox, N., et al.\ 2011, \mnras, 410, 860. doi:10.1111/j.1365-2966.2010.17489.x

\bibitem[\protect\citeauthoryear{Arav et al.}{1999a}]{arav99a} Arav N., Korista K.~T., de Kool M., Junkkarinen V.~T., Begelman M.~C., 1999a, ApJ, 516, 27. doi:10.1086/307073
\bibitem[\protect\citeauthoryear{Arav et al.}{1999b}]{arav99b} Arav N., Becker R.~H., Laurent-Muehleisen S.~A., Gregg M.~D., White R.~L., Brotherton M.~S., de Kool M., 1999b, ApJ, 524, 566. doi:10.1086/307841
\bibitem[\protect\citeauthoryear{Arav et al.}{2001}]{arav01} Arav N., Brotherton M.~S., Becker R.~H., Gregg M.~D., White R.~L., Price T., Hack W., 2001, ApJ, 546, 140. doi:10.1086/318244
\bibitem[\protect\citeauthoryear{Arav et al.}{2005}]{arav05} Arav N., Kaastra J., Kriss G.~A., Korista K.~T., Gabel J., Proga D., 2005, ApJ, 620, 665. doi:10.1086/425560
\bibitem[\protect\citeauthoryear{Arav et al.}{2013}]{arav13} Arav N., Borguet B., Chamberlain C., Edmonds D., Danforth C., 2013, MNRAS, 436, 3286. doi:10.1093/mnras/stt1812
\bibitem[\protect\citeauthoryear{Arav et al.}{2018}]{arav18}Arav, N., Liu, G., Xu, X., et al. 2018, ApJ,857, 60

\bibitem[\protect\citeauthoryear{Astropy Collaboration et al.}{2013}]{astro13} Astropy Collaboration, Robitaille T.~P., Tollerud E.~J., Greenfield P., Droettboom M., Bray E., Aldcroft T., et al., 2013, A\&A, 558, A33. doi:10.1051/0004-6361/201322068
\bibitem[\protect\citeauthoryear{Astropy Collaboration et al.}{2018}]{astro18} Astropy Collaboration, Price-Whelan A.~M., Sip{\H{o}}cz B.~M., G{\"u}nther H.~M., Lim P.~L., Crawford S.~M., Conseil S., et al., 2018, AJ, 156, 123. doi:10.3847/1538-3881/aabc4f
\bibitem[\protect\citeauthoryear{Barlow et al.}{1997}]{barl97} Barlow T.~A., Hamann F., Sargent W.~L.~W., 1997, ASPC, 128, 13. doi:10.48550/arXiv.astro-ph/9705048
\bibitem[\protect\citeauthoryear{Ballero et al.}{2008}]{ball08} Ballero S.~K., Matteucci F., Ciotti L., Calura F., Padovani P., 2008, A\&A, 478, 335. doi:10.1051/0004-6361:20078663
\bibitem[\protect\citeauthoryear{Borguet et al.}{2012}]{borg12b} Borguet B.~C.~J., Edmonds D., Arav N., Benn C., Chamberlain C., 2012, ApJ, 758, 69. doi:10.1088/0004-637X/758/1/69
 \bibitem[Borguet et al.(2013)]{borg13} Borguet, B.~C.~J., Arav, N., Edmonds, D., et al.\ 2013, \apj,  762, 1, 49. doi:10.1088/0004-637X/762/1/49
 \bibitem[Bowler et al.(2014)]{bow14} Bowler, R.~A.~A., Hewett, P.~C., Allen, J.~T., et al.\ 2014, \mnras, Line-driven radiative outflows in luminous quasars, 445, 1, 359. doi:10.1093/mnras/stu1730
\bibitem[\protect\citeauthoryear{Byun et. al}{2022a}]{byun22a} Byun D., Arav N., Walker A., 2022a, MNRAS, 516, 100. doi:10.1093/mnras/stac2194
\bibitem[\protect\citeauthoryear{Byun et. al}{2022b}]{byun22b} Byun D., Arav N., Hall P.~B., 2022b, MNRAS, 517, 1048. doi:10.1093/mnras/stac2638
\bibitem[\protect\citeauthoryear{Byun et. al}{2022c}]{byun22c} Byun D., Arav N., Hall P.~B., 2022c, ApJ, 927, 176. doi:10.3847/1538-4357/ac503d
\bibitem[\protect\citeauthoryear{Chaussidon et al.}{2023}]{chas23} Chaussidon E., Y{\`e}che C., Palanque-Delabrouille N., Alexander D.~M., Yang J., Ahlen S., Bailey S., et al., 2023, ApJ, 944, 107. doi:10.3847/1538-4357/acb3c2

\bibitem[Choi et al.(2022)]{choi22} Choi, H., Leighly, K.~M., Terndrup, D.~M., et al.\ 2022, \apj, 937, 74. doi:10.3847/1538-4357/ac61d9
\bibitem[Churchill et al.(1999)]{chur99} Churchill, C.~W., Schneider, D.~P., Schmidt, M., et al.\ 1999, \aj, 117, 2573. doi:10.1086/300884
\bibitem[\protect\citeauthoryear{Chen et al.}{2022}]{chen22} Chen Z., He Z., Ho L.~C., Gu Q., Wang T., Zhuang M., Liu G., et al., 2022, NatAs, 6, 339. doi:10.1038/s41550-021-01561-3
\bibitem[Coatman et al.(2016)]{coat16} Coatman, L., Hewett, P.~C., Banerji, M., et al.\ 2016, \mnras, 461, 647. doi:10.1093/mnras/stw1360
\bibitem[\protect\citeauthoryear{Coatman et al.}{2017}]{coat17} Coatman L., Hewett P.~C., Banerji M., Richards G.~T., Hennawi J.~F., Prochaska J.~X., 2017, MNRAS, 465, 2120. doi:10.1093/mnras/stw2797
\bibitem[Dai et al.(2012)]{dai12} Dai, X., Shankar, F., \& Sivakoff, G.~R.\ 2012, \apj, 757, 180. doi:10.1088/0004-637X/757/2/180
\bibitem[Davidson \& Netzer (1979)]{dav79}Davidson, K., \& Netzer, H. 1979, Reviews of Modern Physics, 51,715
\bibitem[Dehghanian et al.(2024a)]{deh24} Dehghanian, M., Arav, N., Byun, D., et al.\ 2024a, \mnras, 527, 7825. doi:10.1093/mnras/stad3695
\bibitem[Dehghanian et al.(2024b)]{deh24storm} Dehghanian, M., Arav, N., Kriss, G.~A., et al.\ 2024b, \apj, 972, 141. doi:10.3847/1538-4357/ad5ff4
\bibitem[Dehghanian et al.(2025a)]{deh24b} Dehghanian, M., Arav, N., Sharma, M., et al. 2025a, \aap, 693, A153. doi:10.1051/0004-6361/202452115
\bibitem[Dehghanian et al.(2025b)]{deh25b} Dehghanian, M., Arav, N., Sharma, M., et al.\ 2025b, \aap, 695, A4. doi:10.1051/0004-6361/202453384
\bibitem[\protect\citeauthoryear{Dere et al.}{1997}]{dere97} Dere K.~P., Landi E., Mason H.~E., Monsignori Fossi B.~C., Young P.~R., 1997, A\&AS, 125, 149. doi:10.1051/aas:1997368
\bibitem[\protect\citeauthoryear{Dere et al.}{2019}]{dere19} Dere K.~P., Del Zanna G., Young P.~R., Landi E., Sutherland R.~S., 2019, ApJS, 241, 22. doi:10.3847/1538-4365/ab05cf

\bibitem[DESI Collaboration et al.(2024)]{DESI} DESI Collaboration, Adame, A.~G., Aguilar, J., et al.\ 2024, \aj, 168, 58. doi:10.3847/1538-3881/ad3217
\bibitem[\protect\citeauthoryear{de Kool et. al}{2002}]{kool02} de Kool M., Korista K.~T., Arav N., 2002, ApJ, 580, 54. doi:10.1086/343107
\bibitem[\protect\citeauthoryear{Dunn et al.}{2010}]{dun10} Dunn J.~P., Bautista M., Arav N., Moe M., Korista K., Costantini E., Benn C., et al., 2010, ApJ, 709, 611. doi:10.1088/0004-637X/709/2/611
\bibitem[Filbert et al.(2024)]{fil24} Filbert, S., Martini, P., Seebaluck, K., et al.\ 2024, \mnras, 532, 3669. doi:10.1093/mnras/stae1610

\bibitem[\protect\citeauthoryear{Gabel et al.}{2003}]{gabel03} Gabel J.~R., Crenshaw D.~M., Kraemer S.~B., Brandt W.~N., George I.~M., Hamann F.~W., Kaiser M.~E., et al., 2003, ApJ, 583, 178. doi:10.1086/345096
\bibitem[Gabel et al.(2005)]{gabel05} Gabel, J.~R., Kraemer, S.~B., Crenshaw, D.~M., et al.\ 2005, \apj, 631, 2, 741. doi:10.1086/432682
\bibitem[Gibson et al.(2009)]{gib09} Gibson, R.~R., Jiang, L., Brandt, W.~N., et al.\ 2009, \apj, 692, 758. doi:10.1088/0004-637X/692/1/758
\bibitem[Gunasekera et al.(2023)]{cloud23} Gunasekera, C.~M., van Hoof, P.~A.~M., Chatzikos, M., et al.\ 2023, Research Notes of the American Astronomical Society, 7, 246. doi:10.3847/2515-5172/ad0e75

\bibitem[Ganguly \& Brotherton(2008)]{gang08} Ganguly, R. \& Brotherton, M.~S.\ 2008, \apj, 672, 102. doi:10.1086/524106
\bibitem[\protect\citeauthoryear{Grevesse \& Sauval}{1998}]{grev98} Grevesse N., Sauval A.~J., 1998, SSRv, 85, 161. doi:10.1023/A:1005161325181
\bibitem[Hamann \& Sabra(2004)]{ham04} Hamann, F. \& Sabra, B.\ 2004, AGN Physics with the Sloan Digital Sky Survey, 311, 203. doi:10.48550/arXiv.astro-ph/0310668
\bibitem[Hamann et al.(2013)]{ham13} Hamann, F., Chartas, G., McGraw, S., et al.\ 2013, \mnras, 435, 133. doi:10.1093/mnras/stt1231
\bibitem[Hamann et al.(2019)]{ham19} Hamann, F., Herbst, H., Paris, I., et al.\ 2019, \mnras, 483, 1808. doi:10.1093/mnras/sty2900
\bibitem[\protect\citeauthoryear{Harris et al.}{2020}]{harr20} Harris C.~R., Millman K.~J., van der Walt S.~J., Gommers R., Virtanen P., Cournapeau D., Wieser E., et al., 2020, Natur, 585, 357. doi:10.1038/s41586-020-2649-2
\bibitem[\protect\citeauthoryear{He et al.}{2022}]{he22} 
He Z., Liu G., Wang T., et al. 2022, Science Advances, 8, eabk3291
\bibitem[Hewett \& Foltz(2003)]{hew03} Hewett, P.~C. \& Foltz, C.~B.\ 2003, \aj, 125, 1784. doi:10.1086/368392
\bibitem[\protect\citeauthoryear{Holweger}{2001}]{hol01} Holweger H., 2001, AIPC, 598, 23. doi:10.1063/1.1433974
\bibitem[\protect\citeauthoryear{Hopkins \& Elvis}{2010}]{hop10} Hopkins P.~F., Elvis M., 2010, MNRAS, 401, 7. doi:10.1111/j.1365-2966.2009.15643.x
\bibitem[\protect\citeauthoryear{Hunter}{2007}]{hunt07} Hunter J.~D., 2007, CSE, 9, 90. doi:10.1109/MCSE.2007.55

\bibitem[\protect\citeauthoryear{Itoh et al.}{2020}]{itoh20} Itoh D., Misawa T., Horiuchi T., Aoki K., 2020, MNRAS, 499, 3094. doi:10.1093/mnras/staa2793
\bibitem[King \& Pounds(2015)]{king15} King, A. \& Pounds, K.\ 2015, \araa, 53, 115. doi:10.1146/annurev-astro-082214-122316
\bibitem[Korista et al.(2008)]{kor08} Korista, K.~T., Bautista, M.~A., Arav, N., et al.\ 2008, \apj, Physical Conditions in Quasar Outflows: Very Large Telescope Observations of QSO 2359-1241, 688, 1, 108. doi:10.1086/592140
\bibitem[\protect\citeauthoryear{Krolik }{1999}]{krol99}Krolik, J. H. 1999, Active galactic nuclei: from the central black
hole to the galactic environment, Vol. 59 (Princeton University
Press)
\bibitem[\protect\citeauthoryear{Mathews \& Ferland}{1987}]{mat87} Mathews W.~G., Ferland G.~J., 1987, ApJ, 323, 456. doi:10.1086/165843

\bibitem[\protect\citeauthoryear{Miller et al.}{2020}]{mill20a} Miller T.~R., Arav N., Xu X., Kriss G.~A., Plesha R.~J., 2020, ApJS, 247, 39. doi:10.3847/1538-4365/ab5967
\bibitem[Misawa et al.(2007a)]{misa07} Misawa, T., Charlton, J.~C., Eracleous, M., et al.\ 2007a, \apjs, 171, 1. doi:10.1086/513713

\bibitem[Misawa et al. (2007b)]{misa07b} Misawa T., Eracleous M., Charlton J.~C., Ganguly R., Tytler D., Kirkman D., Suzuki N., et al., 2007b, ASPC, 373, 291. doi:10.48550/arXiv.astro-ph/0612028
\bibitem[\protect\citeauthoryear{Misawa et al.}{2010}]{misa10} Misawa T., Kawabata K.~S., Eracleous M., Charlton J.~C., Kashikawa N., 2010, ApJ, 719, 1890. doi:10.1088/0004-637X/719/2/1890
\bibitem[Napolitano et al.(2023)]{nap23} Napolitano, L., Pandey, A., Myers, A.~D., et al.\ 2023, \aj, 166, 99. doi:10.3847/1538-3881/ace62c

\bibitem[\protect\citeauthoryear{Osterbrock \& Ferland}{2006}]{oster06} Osterbrock D.~E., Ferland G.~J., 2006, agna.book
\bibitem[\protect\citeauthoryear{Prochaska et al.}{2017}]{pro17} Prochaska J.~X., Werk J.~K., Worseck G., Tripp T.~M., Tumlinson J., Burchett J.~N., Fox A.~J., et al., 2017, ApJ, 837, 169. doi:10.3847/1538-4357/aa6007

\bibitem[Planck Collaboration et al.(2020)]{plan20} Planck Collaboration, Aghanim, N., Akrami, Y., et al.\ 2020, \aap, 641, A6. doi:10.1051/0004-6361/201833910
\bibitem[\protect\citeauthoryear{Reback et al.}{2021}]{reba21} Reback J., McKinney W., jbrockmendel, Van den Bossche J., Augspurger T., Cloud P., Hawkins S., et al., 2021, zndo
\bibitem[\protect\citeauthoryear{Redfield \& Falcon}{2008}]{red08} Redfield S., Falcon R.~E., 2008, ApJ, 683, 207. doi:10.1086/589230
\bibitem[\protect\citeauthoryear{Richards et al.}{2011}]{rich11} Richards G.~T., Kruczek N.~E., Gallagher S.~C., Hall P.~B., Hewett P.~C., Leighly K.~M., Deo R.~P., et al., 2011, AJ, 141, 167. doi:10.1088/0004-6256/141/5/167

\bibitem[\protect\citeauthoryear{Savage \& Sembach}{1991}]{sava91} Savage B.~D., Sembach K.~R., 1991, ApJ, 379, 245. doi:10.1086/170498
\bibitem[\protect\citeauthoryear{Scannapieco \& Oh}{2004}]{scan04} Scannapieco E., Oh S.~P., 2004, ApJ, 608, 62. doi:10.1086/386542

\bibitem[\protect\citeauthoryear{Sharma el al.}{2024}]{mayank24} Sharma M., Arav N., Zhao Q.,Dehghanian M., et al. 2024, submitted to ApJ
 \bibitem[Sharma et al.(2025)]{sharma25} Sharma, M., Arav, N., Korista, K.~T., et al.\ 2025, \aap, 693, A254. doi:10.1051/0004-6361/202452735
\bibitem[\protect\citeauthoryear{Silk \& Rees}{1998}]{silk98} Silk J., Rees M.~J., 1998, A\&A, 331, L1. doi:10.48550/arXiv.astro-ph/9801013

\bibitem[Somerville \& Dav{\'e}(2015)]{some15} Somerville, R.~S. \& Dav{\'e}, R.\ 2015, \araa, 53, 51. doi:10.1146/annurev-astro-082812-140951
\bibitem[\protect\citeauthoryear{Spitzer}{1968}]{spit68} Spitzer L., 1968, dms..book
\bibitem[Trump et al.(2006)]{trump06} Trump, J.~R., Hall, P.~B., Reichard, T.~A., et al.\ 2006, \apjs, 165, 1. doi:10.1086/503834
\bibitem[\protect\citeauthoryear{Vayner et al.}{2021a}]{vayner21} Vayner A., Wright S.~A., Murray N., Armus L., Boehle A., Cosens M., Larkin J.~E., et al., 2021, ApJ, 919, 122. doi:10.3847/1538-4357/ac0f56
\bibitem[Vayner et al.(2021b)]{vay21} Vayner, A., Zakamska, N., Wright, S.~A., et al.\ 2021, \apj, 923, 59. doi:10.3847/1538-4357/ac2b9e

\bibitem[\protect\citeauthoryear{Vestergaard}{2003}]{vest03} Vestergaard M., 2003, ApJ, 599, 116. doi:10.1086/379159
\bibitem[\protect\citeauthoryear{Vestergaard \& Peterson}{2006}]{vest06} Vestergaard M., Peterson B.~M., 2006, ApJ, 641, 689. doi:10.1086/500572

\bibitem[\protect\citeauthoryear{Virtanen et al.}{2020}]{virt20} Virtanen P., Gommers R., Oliphant T.~E., Haberland M., Reddy T., Cournapeau D., Burovski E., et al., 2020, NatMe, 17, 261. doi:10.1038/s41592-019-0686-2
 \bibitem[\protect\citeauthoryear{Walker et. al}{2022}]{walk22} Walker A., Arav N., Byun D., 2022, MNRAS, 516, 3778. doi:10.1093/mnras/stac2349
 \bibitem[\protect\citeauthoryear{Weymann et al.}{1991}]{weym91} Weymann R.~J., Morris S.~L., Foltz C.~B., Hewett P.~C., 1991, ApJ, 373, 23. doi:10.1086/170020
\bibitem[\protect\citeauthoryear{Wu \& Shen}{2023}]{wu23} Wu Q., Shen Y., 2023, RNAAS, 7, 190. doi:10.3847/2515-5172/acf580

\bibitem[Xu et al.(2019)]{xu19} Xu, X., Arav, N., Miller, T., et al.\ 2019, \apj, 876, 105. doi:10.3847/1538-4357/ab164e

\bibitem[\protect\citeauthoryear{Yuan et al.}{2018}]{yuan18} Yuan F., Yoon D., Li Y.-P., Gan Z.-M., Ho L.~C., Guo F., 2018, ApJ, 857, 121. doi:10.3847/1538-4357/aab8f8
\end{thebibliography}





\bsp	
\label{lastpage}
\end{document}